# A Hunt for Ultrahard Materials

Vladimir L. Solozhenko [a,*] and Yann Le Godec [b]

[a] *LSPM–CNRS, Université Paris Nord, 93430 Villetaneuse, France*

[b] *Institut de Minéralogie, de Physique des Matériaux et de Cosmochimie (IMPMC), Sorbonne Université, UMR CNRS 7590, Muséum National d'Histoire Naturelle, IRD UMR 206, 75005 Paris, France*

Superhard materials (i.e. those having a load-invariant Vickers hardness $H_V \geq 40$ GPa) are widely used in many industrial applications (abrasives, cutting tool materials, wear-resistant coatings, etc.) where high hardness and shear strength, large elastic moduli, high melting temperatures and chemical inertness are of crucial importance [1].

Vickers hardness of the vast majority of superhard materials is in the range of 40-60 GPa, i.e. does not exceed the hardness of single-crystal cubic boron nitride (for details, see Ref. 1 and references therein). Until very recently diamond was the only known material which is ultrahard. Diamond's unique properties such as extreme hardness (up to 120 GPa [2]), high thermal conductivity, wide band gap, high electron and hole mobility make it suitable for a variety of scientific and technological applications, however, it is rather reactive with oxygen and ferrous metals. Besides, hardness of diamond drastically drops with temperature, down to ~20 GPa at 1400 K [3]. Thus, growing demand for advanced materials for current and emerging needs stimulated the search for novel ultrahard materials that are more thermally and chemically stable than pure diamond. At the very end of the last century, some materials, e.g. 3D-polymerized fullerites [4] and hypothetical low-compressibility carbon nitrides [5] were claimed to be harder than diamond, but as it has been clearly shown very recently that is definitely not the case [6].

At the turn of the millennium, cubic $BC_2N$, a ternary compound that is halfway between diamond and BN in composition has been synthesized at extreme pressure–temperature conditions [7]. The load-independent Vickers hardness of c-$BC_2N$ is 76(4) GPa [8] which made it the second after diamond hardest phase and a member of the ultrahard materials family. The elastic recovery of *c*-$BC_2N$ is 68% which is higher than that of cubic BN (60%) and approaching that of diamond. At ambient pressure in argon atmosphere c-$BC_2N$ remains stable up to 1800 K, and, hence, is characterized by remarkedly higher thermal stability than polycrystalline diamond with the same grain size. A year later, ultrahard low-compressible ($B = 420$ GPa) cubic BN-C solid solutions with stoichiometry close to BCN and statistically uniform atom distribution were synthesized by isoentropic shock compression [9].

---

[*] vladimir.solozhenko@univ-paris13.fr

In 2009 another ultrahard phase, diamond-like $BC_5$, was synthesized under high pressure – high temperature conditions [10]. This new phase corresponds to the ultimate solubility of boron in diamond and possesses Vickers hardness of 71(8) GPa, unusually high for superhard materials fracture toughness (~10 MPa·m$^{1/2}$), and very high (up to 1900 K) thermal stability which makes it an exceptional superabrasive overcoming diamond and promising material for high-temperature electronics. The beneficial combination of electrical conductivity, band structure that is uncommon for diamond-like phases due to electron deficiency of boron atoms, and high thermal stability will eventually allow the expansion of the boundaries of high-power electronics at extreme conditions.

In should be noted that diamond-like $BC_5$ and cubic $BC_2N$ are metastable phases and can be synthesized only in the relatively narrow temperature ranges under pressures of ~20 GPa which makes their production quite a challenge. Superhard and ultrahard ternary B–C–N structures of other stoichiometries ($BC_6N$ [11], $BC_8N$ [12], etc.) have been predicted from first-principles studies, but none of them have ever been synthesized.

The ultrahardness can be also achieved extrinsically by creation of nanostructures and/or nanocomposites that leads to significant increase of the material hardness, mainly due to the grain-boundary strengthening (Hall–Petch effect) (see Ref. 6 and references therein).

In 2003 synthesis of nanocrystalline diamond by direct solid-phase transformation of graphite at pressures above 12 GPa and high (2600-2800 K) temperatures was reported [13]. Knoop hardness of such nanocrystalline bulks reached 140 GPa, which is significantly higher than that of single-crystal diamond. Aggregated diamond nanorods synthesized by direct transition of $C_{60}$ fullerite at 20 GPa and 2500 K were claimed to be the densest and least compressible form of carbon [14]. Nanotwinned diamond with claimed Vickers hardness up to 200 GPa was recently synthesized at 20 GPa and 2200 K from onion carbon nanoparticles [15], however, such unprecedented hardness should be considered highly questionable [6]. In general, nano-polycrystalline diamond materials (irrespective of the precursor, i.e. graphite, fullerenes, carbon nanotubes, soot, etc.) are ultrahard and have significantly higher wear resistance and fracture toughness than single-crystal diamond. In contrast, Vickers hardness of highly polymerized 3D fullerites does not exceed 65 GPa [16].

The first attempt to synthesize nanocrystalline cBN resulted in the formation of ultrahard aggregated nanocomposite of cubic and wurtzitic (wBN) polymorphs [17]. This nanostructured material has Vickers hardness above 80 GPa, but its thermal stability is relatively low due to the presence of metastable wBN. Successful synthesis of single-phase nanocrystalline cBN was performed by direct solid-state phase transformation of graphite-like BN with "ideal random layer" structure at 20 GPa and 1770 K [18]. The material shows very high hardness ($H_V$ = 85(3) GPa) and superior fracture toughness ($K_{Ic}$ = 10.5 MPa·m$^{1/2}$) [19], as well as high thermal stability and oxidation resistance (up to 1500 K) [18]. Later synthesis of ultrahard nanotwinned cBN has been reported [20], however, extremely high Vickers hardness (up to 108 GPa) claimed by the authors is unjustified [21].

Another concept for the design of novel ultrahard materials involves the formation of nanocomposites consisting of two or more nanocrystalline hard phases (e.g. TiN, VN, $W_2N$, etc.) imbedded into amorphous matrix (e.g. $Si_3N_4$, BN, etc.) by chemical vapor deposition (CVD), reactive sputtering or vacuum arc evaporation combined with plasma CVD [22,23]. In particular, nanocrystalline TiN/$TiSi_x$ coatings with Vickers hardness from 80 to 105 GPa have been reported [23], however, these values seem to be overestimated.

In the last two decades, along with experimental work on synthesis of ultrahard materials, theoretical methods for predicting the mechanical properties of solids have been developed very rapidly, from empirical, but physically motivated models of hardness and fracture toughness [24-27] to *ab-initio* calculations of elastic constants [28,29] and computational discovery of superhard and ultrahard materials [30]. The proposed methods work successfully even in the case of boron-rich solids which are characterized by an extreme complexity of the crystal structure and a large number of atoms in a unit cell [31,32]. Thus, the modern evolutionary algorithms and powerful computer systems can be used to predict new structures and compounds, and to search for materials with optimal mechanical properties in order to create a "treasure map" of superhard materials. At the same time, precise calculations of mechanical properties of superhard materials (hardness, in particular) often lie beyond the capabilities of the most advanced and modern techniques. Besides, not all theoretically predicted structures exist or can be synthesized. A vivid illustration is the case of hypothetical cubic form of $C_3N_4$ with bulk modulus exceeding that of diamond [5]. Despite the enormous efforts (new attempts are still being undertaken [33]), this phase has not been synthesized so far, and its expected ultrahardness has never been demonstrated.

To summarize, we should note that the search for novel ultrahard materials is just at the cutting edge of fundamental science and promises great prospects for the creation of new technologies required for emerging applications. In particular, over the past decade nanodiamonds have progressed to fully functioning materials that can be used for a wide range of applications. Recent achievements in the field clearly indicate that synthesis of phases with hardness exceeding that of diamond is very unlikely (or even impossible [6]). Rather than harder, we have to consider the possibility to synthesize (or design) materials that are more useful that diamond i.e. more thermally and chemically stable, and harder than cubic boron nitride.

ORCID IDs : Vladimir L. Solozhenko  https://orcid.org/0000-0002-0881-9761

**References**

1. Y. Le Godec, A. Courac, and V. L. Solozhenko, *J. Appl. Phys.* **126**, 151102 (2019).
2. A. N. Sokolov, A. A. Shul'zhenko, V. G. Gargin *et al.*, *J. Superhard Mater.* **34**, 166-172 (2012).
3. V. V. Brazhkin, and A. G. Lyapin, in *Proc. of NATO Advanced Research Workshop on Innovative Superhard Materials and Sustainable Coatings* (Eds. J. Lee and N. Novikov, Springer, 2005).
4. V. D. Blank, S. G. Buga, N. R Serebryanaya *et al.*, *Carbon* **36**, 665-670 (1998)
5. D. M. Teter, and R. J. Hemley, *Science* **271**, 53-55 (1996).
6. V. V. Brazhkin, and V. L. Solozhenko, *J. Appl. Phys.* **125**, 130901 (2019).
7. V.L. Solozhenko, D. Andrault, G. Fiquet *et al.*, *Appl. Phys. Lett.* **78**, 1385-1387 (2001).
8. V. L. Solozhenko, S. N. Dub, and N. V. Novikov, *Diam. Relat. Mater.* **10**, 2228-2231 (2001).
9. V. L. Solozhenko, *High Press. Res.* **22**, 519-524 (2002).
10. V. L. Solozhenko, O. O. Kurakevych, D. Andrault *et al.*, *Phys. Rev. Lett.* **102**, 015506 (2009).
11. X. Luo, X. Guo, Z. Liu *et al.*, *J. Appl. Phys.* **101**, 083505 (2007).
12. Y. Gao, P. Ying, Y. Wu *et al.*, *J. Appl. Phys.* **125**, 175108 (2019).
13. T. Irifune, A. Kurio, S. Sakamoto *et al.*, *Nature*, **421**, 599-600 (2003).
14. N. Dubrovinskaia, L. Dubrovinsky, W. Crichton *et al.*, *Appl. Phys. Lett.*, **87**, 083106 (2005).
15. Q. Huang, D. Yu, B. Xu *et al.*, *Nature*, **510**, 250 (2014).
16. A. G. Lyapin, Y. Katayama, and V. V. Brazhkin, *J. Appl. Phys.* **126**, 065102 (2019).
17. N. Dubrovinskaya, V. L. Solozhenko, N. Miyajima *et al.*, *Appl. Phys. Lett.* **90**, 101912 (2007).
18. V. L. Solozhenko, O. O. Kurakevych, and Y. Le Godec, *Adv. Mater.* **24**, 1540-1544 (2012).
19. V. L. Solozhenko, V. Bushlya, and J. Zhou, *J. Appl. Phys.* **126**, 075107 (2019).
20. Y. Tian, B. Xu, D. Yu *et al.*, *Nature* **493**, 385-388 (2013).
21. N. Dubrovinskaia, and L. Dubrovinsky, *Nature*, **502**, E1 (2013).
22. S. Veprěk, S. Reiprich, and L. Shizhi, *Appl. Phys. Lett.* **66**, 2640-2642 (1995).
23. P. Nesládek, and S. Veprěk, *Phys. Stat. Sol. (a)* **177**, 53-62 (2000).
24. V. A. Mukhanov, O. O. Kurakevych, and V. L. Solozhenko, *High Press. Res.* **28**, 531-537 (2008).
25. V. A. Mukhanov, O. O. Kurakevych, and V. L. Solozhenko, *Philos. Mag.* **89**, 2117-2127 (2009).
26. E. Mazhnik, and A. R. Oganov, *J. Appl. Phys.* **126**, 125109 (2019).
27. H. Niu, S. Niu, and A. R. Oganov, *J. Appl. Phys.* **125**, 065105 (2019).
28. C. Xie, Q. Zhang, H. A. Zakaryan *et al.*, *J. Appl. Phys.* **125**, 205109 (2019).
29. V. I. Ivashchenko, P. E. A. Turchi, L. Gorb *et al.*, *J. Appl. Phys.* **125**, 075303 (2019).
30. A. G. Kvashnin, Z. Allahyari, and A. R. Oganov, *J. Appl. Phys.* **126**, 040901 (2019).
31. V. A. Mukhanov, O. O. Kurakevych, and V. L. Solozhenko, *J. Superhard Mater.* **32**, 167-176 (2010).
32. A. Jay, O. H. Duparc, J. Sjakste, and N. Vast, *J. Appl. Phys.* **125**, 185902 (2019).
33. X. Gao, H. Yin, P. Chen, and J. Liu, *J. Appl. Phys.* **126**, 155901 (2019).